\documentclass[conference]{IEEEtran}
\usepackage{algorithm,algorithmic}
\makeatletter
\newcommand\fs@betterruled{%
  \def\@fs@cfont{\bfseries}\let\@fs@capt\floatc@ruled
  \def\@fs@pre{\vspace*{5pt}\hrule height.8pt depth0pt \kern2pt}%
  \def\@fs@post{\kern2pt\hrule\relax}%
  \def\@fs@mid{\kern2pt\hrule\kern2pt}%
  \let\@fs@iftopcapt\iftrue}
\floatstyle{betterruled}
\restylefloat{algorithm}

\makeatother

\usepackage{cite}
\usepackage{amsmath,amssymb,amsfonts}
\usepackage{algorithmic}
\usepackage{graphicx}
\usepackage{textcomp}
\usepackage{comment}
\usepackage{enumitem, subcaption}
\usepackage{xcolor}
\def\BibTeX{{\rm B\kern-.05em{\sc i\kern-.025em b}\kern-.08em
    T\kern-.1667em\lower.7ex\hbox{E}\kern-.125emX}}
\renewcommand{\a}{\mathbf{a}}

\newcommand{\n}{\mathbf{n}}

\newcommand{\s}{\mathbf{s}}

\newcommand{\x}{\mathbf{x}}
\newcommand{\y}{\mathbf{y}}

\newcommand{\B}{\mathbf{B}}

\newcommand{\D}{\mathbf{D}}
\newcommand{\E}{\mathbf{E}}
\newcommand{\F}{\mathbf{F}}
\newcommand{\G}{\mathbf{G}}
\renewcommand{\H}{\mathbf{H}}

\newcommand{\K}{\mathbf{K}}

\newcommand{\M}{\mathbf{M}}

\renewcommand{\P}{\mathbf{P}}

\newcommand{\R}{\mathbf{R}}

\newcommand{\U}{\mathbf{U}}
\newcommand{\V}{\mathbf{V}}
\newcommand{\W}{\mathbf{W}}

\newcommand{\Y}{\mathbf{Y}}

\newcommand{\Compl}{\mbox{$\mathbb{C}$}}

\usepackage{xcolor, soul}
\sethlcolor{yellow}
\begin{document}
%\title{Full Duplex Integrated Sensing and Communication for TeraHertz Frequencies with Dynamic Metasurface Antennas}
%\title{Near-Field Integrated Sensing and Communications at THz with Full Duplex Metasurface-Based Antennas} 
\title{Full Duplex Holographic MIMO for\\ Near-Field Integrated Sensing and Communications} 
\author{\IEEEauthorblockN{Ioannis Gavras\IEEEauthorrefmark{1}, Md Atiqul Islam\IEEEauthorrefmark{2}, Besma Smida\IEEEauthorrefmark{3}, and George C. Alexandropoulos\IEEEauthorrefmark{1}}
\IEEEauthorblockA{\IEEEauthorrefmark{1}Department of Informatics and Telecommunications, National and Kapodistrian University of Athens, Greece\\{\IEEEauthorrefmark{2}Qualcomm Technologies, Inc., Santa Clara, CA, USA}\\{\IEEEauthorrefmark{3}Department of Electrical and Computer Engineering, University of Illinois at Chicago, USA}\\
emails: \{sdi1900029, alexandg\}@di.uoa.gr, mdatiqul@qti.qualcomm.com, smida@uic.edu 
}}
\maketitle
\begin{abstract}
This paper presents an in-band Full Duplex (FD) integrated sensing and communications system comprising a holographic Multiple-Input Multiple-Output (MIMO) base station, which is capable to simultaneously communicate with multiple users in the downlink direction, while sensing targets being randomly distributed within its coverage area. Considering near-field wireless operation at THz frequencies, the FD node adopts dynamic metasurface antenna panels for both transmission and reception, which consist of massive numbers of sub-wavelength-spaced metamaterials, enabling reduced cost and power consumption analog precoding and combining. We devise an optimization framework for the FD node’s reconfigurable parameters with the dual objective of maximizing the targets’ parameters estimation accuracy and the downlink communication performance. Our simulation results verify the integrated sensing and communications capability of the proposed FD holographic MIMO system, showcasing the interplays among its various design parameters. 
\end{abstract}


\section{Introduction}
The combination of sensing and communication signaling operations under the same system infrastructure is recently gaining remarkable ground as an efficient means for improving spectral and energy efficiencies in $6$th Generation (6G) networks~\cite{liu2022integrated,mishra2019toward}. This notion of Integrated Sensing and Communications (ISAC) is lately being investigated in the framework of in-band Full Duplex (FD) Multiple-Input Multiple-Output (MIMO) radios, which enable simultaneous DownLink (DL) transmission and uplink reception (of data or control signals) within the entire frequency band \cite{sabharwal2014band,alexandropoulos2017joint,FD_MIMO_VTM2022}.

The main challenge for FD MIMO ISAC systems is the Self-Interference (SI) signal induced from the multi-antenna Transmitter (TX) of the FD node to its multi-antenna Receiver (RX), which increases with the number of TX antenna elements. State-of-the-art solutions include combination of propagation domain isolation, analog domain suppression, digital SI cancellation techniques, and recently, hybrid analog and digital BeamForming (BF), which has been shown to perform efficiently in FD massive MIMO systems operating in millimeter-wave frequencies~\cite{Vishwanath_2020, alexandropoulos2020full}. Single-antenna FD systems realizing joint radar communication and sensing were considered in~ \cite{barneto2019full,liyanaarachchi2021optimized}, while FD ISAC operations with millimeter-wave massive MIMO systems were designed in~\cite{barneto2020beamforming,Islam_2022_ISAC,Atiq_ISAC_2022}, showcasing the efficacy of FD systems for simultaneous data communication and target tracking. However, to the best of the authors' knowledge, FD MIMO ISAC operating in the THz frequency band and in the near-field regime has not yet been reported.

In this paper, we present a novel FD holographic MIMO system, which is realized via the efficiently scalable technology of Dynamic Metasurface Antennas (DMAs) \cite{Shlezinger2021Dynamic} at both its TX and RX ends. Those arrays of metamaterials are designed to enable simultaneous multi-user DL data communication and $3$D direction of arrival as well as range estimation of multiple targets lying in the vicinity of the proposed FD-enabled ISAC system. Considering wireless operation in the THz frequency band and in the near-field regime, we assume that the targets reflect the DL signals back to the FD holographic MIMO system, enabling their spatial parameters' tracking. By modeling signal propagation over the metamaterial-based microstrips comprising each DMA as well as their tunable frequency responses, we present a novel optimization framework for the joint desing of the TX/RX DMAs’ analog BF matrices, the TX digital BF matrix, and the digital SI cancellation matrix, having a dual objective that includes the DL rate and the accuracy of the target parameters’ estimation. An extensive waveform simulation at sub-THz frequencies verifies the performance of the proposed FD holographic MIMO ISAC system.

\textit{Notations:} Vectors and matrices are denoted by boldface lowercase and boldface capital letters, respectively. The transpose, Hermitian transpose, and the inverse of $\mathbf{A}$ are denoted by $\mathbf{A}^{\rm T}$, $\mathbf{A}^{\rm H}$, and $\mathbf{A}^{-1}$, respectively, while $\mathbf{I}_{n}$ and $\mathbf{0}_{n}$ ($n\geq2$) are the $n\times n$ identity and zeros' matrices, respectively. $[\mathbf{A}]_{i,j}$ is the $(i,j)$-th element of $\mathbf{A}$, $\|\mathbf{A}\|$ returns $\mathbf{A}$'s Euclidean norm, $|a|$ is the amplitude of a complex scalar $a$, $\mathbb{C}$ is the complex number set, and $\jmath$ is the imaginary unit. $\mathbb{E}\{\cdot\}$ is the expectation operator and $\mathbf{x}\sim\mathcal{CN}(\mathbf{a},\mathbf{A})$ indicates a complex Gaussian random vector with mean $\mathbf{a}$ and covariance matrix $\mathbf{A}$.

\begin{figure}[!t]
%\begin{figure*}[!tpb]
	\begin{center}
	\includegraphics[width=0.9\columnwidth]{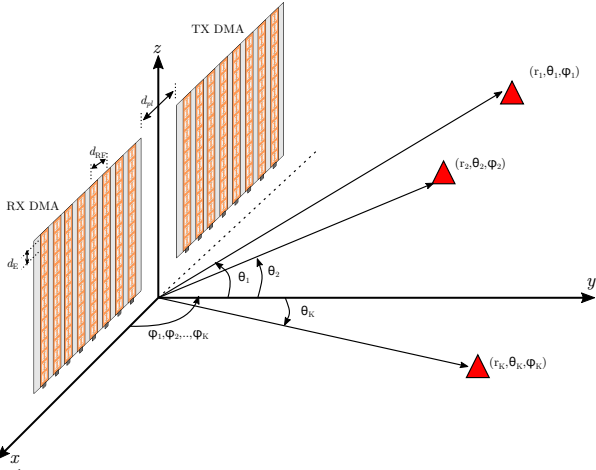}\vspace{-0.2cm}
	\caption{\small{The proposed FD holographic MIMO system enabling ISAC. $d_{\rm RF}$ is the distance between adjacent microstrips in each DMA, $d_{\rm E}$ is the inter-element distance within each microstrip, and $d_{pl}=2d_{\rm P}$ is the horizontal separation between the TX and RX DMAs.}}\vspace{-0.4cm}
	\label{fig: FD_ISAC}
	\end{center}
%\end{figure*}
\end{figure}
\begin{comment}
In this paper, we present a multi-user FD ISAC system including a framework for multiple radar target DoA and range estimation across several communication subframes, operating in sub-terahertz frequencies. The considered ISAC system employs an FD massive MIMO BS node communicating with multiple DL users, and utilizes the reflected waveforms to detect the radar targets residing within the communication environment. We propose a joint design of the A/D beamformers and a reduced complexity SI cancellation for the FD ISAC system, which target at maximizing the multi-user DL communication rate and the precision of the radar target estimation. 
\end{comment}
\section{System and Signal Models}\label{sec: system_signal}
We consider an FD-enabled ISAC system comprising an FD holographic MIMO node equipped with TX and RX DMAs~\cite{Shlezinger2021Dynamic}, which wishes to communicate in the DL direction with $U$ multi-antenna User Equipment (UEs), while simultaneously being capable to sense $K\geq U$ radar targets lying in its vicinity. It is assumed that the $U$ served UEs are fixed, belonging to the $K$ sensed targets. DMAs can efficiently realize holographic MIMO transceivers~\cite{HMIMO_survey}, facilitating packaging larger number of sub-wavelength-spaced metamaterials in given apertures. They consist of single-RF-fed microstrips, where the transmitted/received signals are phase-controlled via metamaterials of dynamically tunable frequency responses, thus, implementing analog BF.  
%The DMAs utilize radiating metamaterial elements embedded onto the surfaces of waveguide microstrips to perform as planar reconfigurable antenna surfaces. The motivation for using DMAs is their capability to pack a larger number of sub-wavelength spaced elements in a given aperture compared to conventional uniform planar arrays resulting in a low-cost and power consumption architecture. The frequency response of each metamaterial element can be controlled by adjusting its local electrical properties. For our FD ISAC architecture, each of the microstrips of the DMAs is fed by a single RF chain, where the input signal is radiated by controlled phase-shifted metamaterial elements resulting in an analog beamforming structure. Therefore, the FD system with DMA formulates a hybrid beamforming architecture.
As shown in Fig.~\ref{fig: FD_ISAC}, the TX and RX DMA panels of the FD MIMO node are placed in the $xz$-plane centered at the origin. $N_{\rm RF}$ microstrips with $N_{\rm E}$ metamaterials of inter-element spacing of $d_{\rm E}$ are assumed in both TX and RX DMAs, each connected to a respective Radio Frequency (RF) chain. Hence, both the TX and RX have $N\triangleq N_{\rm RF}N_{\rm E}$ metamaterials. Finally, all $U$ UEs are equipped with $L$-element all-digital Uniform Linear Arrays (ULA), which are placed for simplicity on the $z$-axis.
%Therefore, the FD ISAC node with DMA consists of $N$ and $M$ metamaterial elements, respectively, where $N=M=N_{\rm RF}\times N_{\rm E}$. In the RX users node, we employ an all-digital Uniform Linear Array (ULA) structure with $L$ antenna elements placed in the $z$-axis, where each antenna is connected to an individual RX RF chain.

We define the $N\times N$ diagonal matrices $\P_{\rm TX}$ and $\P_{\rm RX}$, whose elements model the signal propagation inside the microstrips at the TX and RX DMAs, respectively. The former is defined $\forall$$i=1,\dots,N_{\rm RF}$ and $\forall$$n = 1,\dots,N_{\rm E}$ by \cite{Xu_DMA_2022}:
\begin{align}\label{eq: TX_Sig_Prop}
    [\P_{\rm TX}]_{((i-1)N_{\rm E}+n,(i-1)N_{\rm E}+n)} \triangleq \exp{(-\rho_{i,n}(\alpha_i + \jmath\beta_i))},
\end{align}
where $\alpha_i$ is the waveguide attenuation coefficient, $\beta_i$ is the wavenumber, and $\rho_{i,n}$ denotes the location of the $n$th element in the $i$th microstrip. Similar is the definition for $\P_{\rm RX}$.
%In the DMA-based transmission architecture, the signal fed by the RF chains propagates inside the microstrips before being radiated by the metamaterial elements. To develop an input-output relation of the DMAs, the signal propagation is formulated as the diagonal matrix $\P_{\rm TX}\in \mathbb{C}^{N\times N}$, which is defined by
%\begin{align}\label{eq: TX_Sig_Prop}
%    [\P_{\rm TX}]_{((i-1)N_{\rm E}+n,(i-1)N_{\rm E}+n)} \triangleq \exp{(-\rho_{i,n}(\alpha_i + j\beta_i))},\\
%    \nonumber\forall i=1,\dots,N_{\rm RF},\, n = 1,\dots,N_{\rm E},
%\end{align}
%where $\alpha_i$ is the waveguide attenuation coefficient, $\beta_i$ is the wavenumber, and $\rho_{i,n}$ denotes the location of the $n$th element in the $i$th microstrip/RF chain. In a similar manner, the RX signal at the antenna elements propagates inside the RX DMA before feeding into the RF chain. The RX signal propagation is represented by the diagonal matrix $\P_{\rm RX}\in \mathbb{C}^{M\times M}$ whose elements are defined similarly as \eqref{eq: TX_Sig_Prop}.
Let $w^{\rm TX}_{i,n}$ denote the tunable response (i.e., analog weight) of each $n$th metamaterial of each $i$th microstrip, which is assumed to follow a Lorentzian-constrained phase model and belong to the phase profile codebook $\mathcal{W}$:
%These dynamically controllable weights of the metamaterial elements represents the analog beamforming structure for the communication system. Here, we consider that the set of possible values of $w^{\rm TX}_{i,n}$, denoted $\mathcal{W}$, represents a Lorentzian-constrained phase model of the metamaterial elements frequency response, which is defined as
\begin{align}
    w^{\rm TX}_{i,n} \in \mathcal{W}\triangleq \left\{\frac{\jmath+e^{\jmath\phi}}{2}\Big|\phi\in\left[-\frac{\pi}{2},\frac{\pi}{2}\right]\right\}.
\end{align}
The analog TX BF matrix  $\W_{\rm TX}\in\mathbb{C}^{N\times N_{\rm RF}}$ is given by:
\begin{align}
    [\W_{\rm TX}]_{((i-1)N_{\rm E}+n,j)} = \begin{cases}
    w^{\rm TX}_{i,n},&  i=j,\\
    0,              & i\neq j.
\end{cases}
\end{align}
Similarly, we define the weights $w^{\rm RX}_{i,n} \in \mathcal{W}$ $\forall$$i,n$, from which the analog RX BF matrix  $\W_{\rm RX}\in\mathbb{C}^{N\times N_{\rm RF}}$ is formulated.

The DMA-based TX of the proposed FD holographic MIMO system possesses the symbol vector $\s_{u}\in\Compl^{L\times1}$ for each $u$th UE which is first precoded via the digital BF matrix $\V_u\in\Compl^{N_{\rm RF}\times L}$. Before transmission in the DL, the digitally precoded symbols are analog processed via the weights of the TX DMA, resulting in the $N$-element transmitted signal $\x\triangleq\P_{\rm TX}\W_{\rm TX}\V\s$ where $\V \triangleq [\V_1,\ldots,\V_U]\in\Compl^{N_{\rm RF}\times UL}$ and $\s \triangleq [\s_1^{\rm T},\ldots,\s_U^{\rm T}]^{\rm T}\in\Compl^{UL\times1}$. We finally assume this signal is power limited such that $\mathbb{E}\{\|\P_{\rm TX}\W_{\rm TX}\V\s\|^2\}\leq P_{\rm max}$, where $P_{\rm max}$ being the maximum transmission power.
%\begin{equation}\label{eq:TX_Signal}
%\begin{split}
%   \x &\triangleq \P_{\rm TX}\W_{\rm TX}[\V_1,\ldots,\V_U][\s_1^{\rm T},\ldots,\s_U^{\rm T}]^{\rm T}\\
%& = \P_{\rm TX}\W_{\rm TX}\V\s,
%\end{split}
%\end{equation}
%where $\V \triangleq [\V_1,\ldots,\V_U]\in\Compl^{N_{\rm RF}\times UL}$ and $\s \triangleq [\s_1^{\rm T},\ldots,\s_U^{\rm T}]^{\rm T}\in\Compl^{UL\times1}$.
%\begin{align}
%    \nonumber&\V \triangleq [\V_1,\dots\V_U]\in\Compl^{N_{\rm RF}\times UL},\\
%    &\s \triangleq [\s_1^{\rm T},\dots,\s_U^{\rm T}]^{\rm T}\in\Compl^{UL}.
%\end{align}

\subsection{Near-Field Channel Model}
\begin{comment}
Since large-scale planar DMAs combined with very high transmission frequencies result in operation in the near-field region, we formulate the near-field signaling of the DL and radar transmission taking into account the spherical wavefront at the FD ISAC node. As shown in Fig.~\ref{fig: FD_ISAC}, we consider $K$ radar targets with spherical coordinates $\{(r_1,\theta_1,\varphi_1),\dots,(r_K,\theta_K,\varphi_K)\}$ with respect to the origin distributed in on the $yz$-plane, where $r,\theta$, and $\varphi$ represent the distance, elevation, and azimuth angle, respectively. Here, $U$ out of $K$ targets with spherical coordinates $\{(r_1,\theta_1,\varphi_1),\dots,(r_U,\theta_U,\varphi_U)\}$  are considered as DL users.
\end{comment}
%Because the large-scale planar DMAs are used with very high transmission frequencies, they operate in the near-field region. Therefore, we have formulated the near-field signaling for the radar transmission and DL, taking into account the spherical wavefront at the FD ISAC node. In Fig. 1, we have shown K radar targets with spherical coordinates $\{(r_1,\theta_1,\varphi_1),\dots,(r_K,\theta_K,\varphi_K)\}$ with respect to the origin distributed in on the $yz$-plane, where $r,\theta$, and $\varphi$ represent the distance, elevation, and azimuth angle, respectively. Here, $U$ out of $K$ targets with spherical coordinates $\{(r_1,\theta_1,\varphi_1),\dots,(r_U,\theta_U,\varphi_U)\}$  are considered as DL users.
We consider wireless operation in the THz frequency band, which takes place in a near-field signal propagation environment, as shown in Fig.~\ref{fig: FD_ISAC}. Each $L\times N$ complex-valued DL channel (i.e., for each $u$th UE) is modeled as follows:
\begin{align}
    \label{eqn:DL_chan}
    [\H_{{\rm DL},u}]_{(\ell,(i-1)N_{\rm E}+n)} \triangleq \alpha_{u,\ell,i,n} \exp\left(\frac{\jmath2\pi}{\lambda} r_{u,\ell,i,n}\right),
\end{align}
where $r_{u,\ell,i,n}$ represents the distance between the $\ell$th antenna ($\ell=1,\ldots,L$) of each $u$th UE and the $n$th meta-element of each $i$th TX DMA's RF chain. In addition, $\alpha_{u,\ell,i,n}$ defines the respective attenuation factor with molecular absorption coefficient $\kappa_{\rm abs}$ at THz, which is defined as:
\begin{align}\label{eq: atn}
    \alpha_{u,\ell,i,n} \triangleq \sqrt{F(\theta_{u,\ell,i,n})} \frac{\lambda}{4\pi r_{u,\ell,i,n}} \exp\left(-\frac{\kappa_{\rm abs}r_{u,\ell,i,n}}{2}\right)
\end{align}
with $\lambda$ being the wavelength and $F(\cdot)$ is each metamaterial's radiation profile, modeled for an elevation angle $\theta$ as follows:
\begin{align}
    F (\theta) = \begin{cases}
    2(b+1)\cos^{b}(\theta),& {\rm if}\, \theta\in[-\frac{\pi}{2},\frac{\pi}{2}],\\
    0,              & {\rm otherwise}.
\end{cases}
\end{align}
In the latter expression, $b$ determines the boresight antenna gain which depends on the specific DMA technology. 

As depicted in Fig.~\ref{fig: FD_ISAC}, we consider $K$ targets with spherical coordinates $\{(r_1,\theta_1,\varphi_1),\ldots,(r_K,\theta_K,\varphi_K)\}$, including the distances from the origin, and the elevation and azimuth angles, respectively. From those targets, $U$ out of $K$ with coordinates $\{(r_1,\theta_1,\varphi_1),\dots,(r_U,\theta_U,\varphi_U)\}$ are the DL UEs. Each distance $r_{u,\ell,i,n}$ in \eqref{eqn:DL_chan} and \eqref{eq: atn} can be calculated as:
\begin{align}\label{eq: dist}
    \nonumber &r_{u,\ell,i,n}\! =\! \!\Big(\!(r_{u,\ell}\sin\theta_{u,\ell}\cos\varphi_{u,\ell} +\frac{d_{\rm P}}{2}+(i\!-\!1)d_{\rm RF})^2 +\\ &(r_{u,\ell}\sin\theta_{u,\ell}\sin\varphi_{u,\ell})^2 + (r_{u,\ell}\sin\theta_{u,\ell}\!-\!(n\!-\!1)d_{\rm E})^2\Big)^{\frac{1}{2}},
\end{align}
where $r_{u,\ell}$, $\theta_{u,\ell}$, and $\varphi_{u,\ell}$ represent the distance, and elevation and azimuth angles of each $u$th UE's $\ell$th antenna with respect to the origin, which can be computed as follows:
\begin{align}
    &\nonumber\theta_{u,\ell}=\tan^{-1}\Big(\frac{r_u \sin\theta_u+(\ell\!-\!1)d_{\rm RF}}{r_u \cos\theta_u}\Big),\\
    & r_{u,\ell}=\frac{r_u \cos\theta_u}{\cos\theta_{u,\ell}},
    \quad\varphi_{u,\ell} \triangleq \varphi_u.
\end{align}
Note that the elevation angle of each $u$th UE's $\ell$th antenna with respect to the $n$th element of each $i$th microstrip is:
\begin{align}\label{eq:thetas}
    \theta_{u,\ell,i,n} \triangleq \sin^{-1}\left({\frac{|(n-1)d_{\rm E}-r_{u,\ell}\cos{\theta_{u,\ell}}|}{r_{u,\ell,i,n}}}\right).
\end{align}

The end-to-end channel model including the impinging/reflected components to/from the $K$ targets, when considered as point sources with coordinates $(r_k,\theta_k,\varphi_k)$ $\forall$$k=1,\ldots,K$, can be expressed as follows: 
\begin{align}\label{eq:H_R}
    \H_{\rm R} \triangleq \sum\limits_{k=1}^{K}\beta_k \a_{\rm RX}(r_k,\theta_k,\varphi_k)\a_{\rm TX}^{\rm H}(r_k,\theta_k,\varphi_k)
\end{align}
with $\beta_k$ representing the complex-valued reflection coefficient for each $k$th radar target, whereas, using \eqref{eq: atn} and the string definition ${\rm str}\triangleq\{{\rm TX},{\rm RX}\}$, $\a_{\rm str}(\cdot)$ can be obtained as:
\begin{align}
    \label{eq:response_vec}
    [\a_{\rm str}(r_k,\theta_k,\phi_k)]_{(i-1)N_E+n} \triangleq a_{k,i,n}\exp\Big({\jmath\frac{2\pi}{\lambda}r_{k,i,n}}\Big).
\end{align}
%\begin{align}
%    \label{eq:response_vec}
%    \nonumber[\a_{\rm TX}(r_k,\theta_k,\phi_k)]_{(i-1)N_E+n} \triangleq &a_{k,i,n}(r_{k,i,n},\theta_{k,i,n})\\
%    &\nonumber\times\exp\Big({\jmath\frac{2\pi}{\lambda}r_{k,i,n}}\Big)\\
%    \nonumber[\a_{\rm RX}(r_k,\theta_k,\phi_k)]_{(i-1)N_E+n} \triangleq &a_{k,i,n}(r_{k,i,n},\theta_{k,i,n})\\
%    &\times\exp\Big({-\jmath\frac{2\pi}{\lambda}r_{k,i,n}}\Big)
%\end{align}
In this expression, the elevation angle $\theta_{k,i,n}$ and the distance $r_{k,i,n}$ from the origin for each $k$th target are needed to compute $a_{k,i,n}$. The former value can be obtained similar to \eqref{eq:thetas}, while the latter is given by the following expression:
\begin{align}
    \nonumber r_{k,i,n}\triangleq\Big(\big(r_k\sin{\theta_k}\cos{\phi_k}\pm\frac{d_{\rm P}}{2}\pm(i-1)d_{\rm RF}\big)^2\\+\big(r_k\sin{\theta_k}\sin{\phi_k}\big)^2
    +\big(r_k\cos{\theta_k}-(n-1)d_{\rm E}\big)^2\Big)^{\frac{1}{2}}.
\end{align}
In this expression, the positive sign refers to the transmission vector, while the negative sign indicates the reception vector. 
%The distance between a metamaterial antenna element and the target's body is formulated as
%\begin{align}
%    \nonumber r_{k,i,n}\triangleq\Big(\big(r_k\sin{\theta_k}\cos{\phi_k}+\frac{d_{\rm P}}{2}+(i-1)d_{\rm RF}\big)^2\\+\big(r_k\sin{\theta_k}\sin{\phi_k}\big)^2
%    +\big(r_k\cos{\theta_k}-(n-1)d_{\rm E}\big)^2\Big)^{\frac{1}{2}}\nonumber\\
%\end{align}
%Similarly at the receiving end the distance can be formulated as
%\begin{align}
%    \nonumber r_{k,i,n}\triangleq\Big(\big(r_k\sin{\theta_k}\cos{\phi_k}-\frac{d_{\rm P}}{2}-(i-1)d_{\rm RF}\big)^2\\+\big(r_k\sin{\theta_k}\sin{\phi_k}\big)^2
%    +\big(r_k\cos{\theta_k}-(n-1)d_{\rm E}\big)^2\Big)^{\frac{1}{2}}\nonumber\\
%\end{align}
%and for either case we express the respective elevation angle as follows
%\begin{align}
%    \theta_{k,i,n} =& \sin^{-1}{\frac{|(n-1)d_{\rm E}-r_k\cos{\theta_k|}}{r_{i,n}}}
%\end{align}

\subsection{Received Signal Models}
The baseband received signal $\y_{u}\in\Compl^{L\times1}$ at each $u$th UE can be mathematically expressed as:
\begin{align}
    \y_{u} \triangleq \H_{{\rm DL},u}\P_{\rm TX}\W_{\rm TX}\V\s + \n_u,
\end{align}
where $\n_u\sim\mathcal{CN}(\mathbf{0},\sigma^2_u\mathbf{I}_L)$ denotes the Additive White Gaussian Noise (AWGN) vector. Similarly, the baseband received signal $\y\in\Compl^{N_{\rm RF}\times1}$ at the output of the RX DMA panel of the proposed FD holographic MIMO system is given by:
\begin{align}\label{eq:received}
    \y \triangleq& \W^{\rm H}_{\rm RX}\P^{\rm H}_{\rm RX}\H_{\rm R}\P_{\rm TX}\W_{\rm TX}\V\s \\
    &\nonumber+ (\W^{\rm H}_{\rm RX}\P^{\rm H}_{\rm RX}\H_{\rm SI}\P_{\rm TX}\W_{\rm TX} + \D)\V\s + \W^{\rm H}_{\rm RX}\P^{\rm H}_{\rm RX}\n,
    %\\
    %\nonumber=&\W^{\rm H}_{\rm RX}\P^{\rm H}_{\rm RX}\H_{\rm R}\P_{\rm TX}\W_{\rm TX}\V\s + (\widetilde{\H}_{\rm SI}+\D)\V\s+ \W^{\rm H}_{\rm RX}\P^{\rm H}_{\rm RX}\n,
\end{align}
where $\n\sim\mathcal{CN}(\mathbf{0},\sigma^2\mathbf{I}_{N_{\rm RF}})$ denotes the AWGN vector and $\H_{\rm SI}\in\Compl^{M\times N}$ represents the near-field SI channel, which is defined $\forall$$i,i'=1,\dots,N_{\rm RF}$ and $\forall$$n,n'=1,\dots,N_{\rm E}$ as:
\begin{align}
    &\nonumber[\H_{\rm SI}]_{((i-1)N_E+n,(i'-1)N_E+n')} \triangleq \alpha_{i,i',n,n'}\exp\left(\frac{\jmath2\pi}{\lambda}r_{i,i',n,n'}\right),
\end{align}
where $\theta_{i,i',n,n'} = \sin^{-1}\left(|n'-n|d_{\rm E}/r_{i,i',n,n'}\right)$ and 
\begin{align}
    &\nonumber r_{i,i',n,n'} \triangleq \Big(\big(\frac{d_{\rm P}}{2}+(i'-1)d_{\rm RF}-(-\frac{d_P}{2}-(i-1)d_{\rm RF})\big)^2\\
    &\nonumber+\big((n'-1)d_{\rm E}-(n-1)d_{\rm E}\big)^2\Big)^{\frac{1}{2}}.
\end{align}

It is noted that the term $\widetilde{\H}_{\rm SI}\triangleq \W^{\rm H}_{\rm RX}\P^{\rm H}_{\rm RX}\H_{\rm SI}\P_{\rm TX}\W_{\rm TX}$ in \eqref{eq:received} indicates the residual SI contribution, including the impact of the TX and RX DMAs which will be optimized, together with the digital SI cancellation matrix $\D\in\mathbb{C}^{N_{\rm RF}\times N_{\rm RF}}$, for ISAC in the sequel. Since the considered, in this paper, holographic BF is capable of realizing highly directive beams, we will not use analog SI cancellation units \cite{FD_MIMO_VTM2022}.

\section{Proposed FD-Enabled ISAC Framework}
In this section, we design the parameters of the proposed FD holographic MIMO system for near-field ISAC. In particular, we derive the TX/RX DMAs' analog BF matrices, the TX digital BF matrix, and the digital SI cancellation matrix based on a dual objective function including the DL rate and the accuracy of the target parameters' estimation.

\subsection{Near-Field Targets' Parameters Estimation}\label{ssec:MUSIC}
By using $T$ transmission time slots to construct the reception matrix $\Y\in\Compl^{N_{\rm RF}\times T}$ using \eqref{eq:received}, the sample covariance matrix $\R \triangleq \frac{1}{T} \Y\Y^{\rm H}\in\Compl^{N_{\rm RF}\times N_{\rm RF}}$ can be computed for estimating the range, and elevation and azimuth angles of the $K$ targets via MULtiple SIgnal Classification. Performing $\R$'s eigenvalue decomposition, yields $\R = \U {\rm diag}\{\eta_1,\eta_2,\ldots,\eta_{N_{\rm RF}}\}\U^{\rm H}$ with $\eta_1\geq\eta_2\geq\ldots\geq\eta_{N_{\rm RF}}$ being the matrix eigenvalues in descending order and $\U\in\Compl^{N_{\rm RF}\times N_{\rm RF}}$ including the eigenvectors ($\eta_i$ associates with eigenvector $\mathbf{u}_i\triangleq[\U]_{:,i}$). In fact, $\U$ can be partitioned as $\U=[\U_s|\U_n]$, where $\U_n\in\mathbb{C}^{N_{\rm RF}\times N_{\rm RF}-K}$ and $\U_s\in\mathbb{C}^{N_{\rm RF}\times K}$ contain the noise- and signal-subspace eigenvectors, respectively. Capitalizing on the orthogonality between the latter subspaces, the $3$D MUSIC spectrum can be expressed as a function of the unknown target parameters as follows:
\begin{align}\label{Spec_ROOT}
    \mathcal{M}(r,\theta,\varphi) &\triangleq \left(\M^{\rm H}\U_{K}\U_{K}^{\rm H}\M\right)^{-K}\prod_{k=1}^{K-1}\M^{\rm H}\U_{k}\U_{k}^{\rm H}\M,
    %\frac{\prod_{k=1}^{K-1}\M^{\rm H}\U_{k}\U_{k}^{\rm H}\M}{\left(\M^{\rm H}\U_{K}\U_{K}^{\rm H}\M\right)^K},
\end{align}
%\begin{align}\label{Spec_ROOT}
%    \nonumber\mathcal{M}(r,\theta,\varphi) &\triangleq \frac{\M^{H}\U_{\rm n_{\rm K-1}}\U_{\rm n_{\rm K-1}}^H\M}{\M^{H}\U_{\rm n_{\rm K}}\U_{\rm n_{\rm K}}^H\M}\frac{\M^{H}\U_{\rm n_{\rm K-2}}\U_{\rm n_{\rm K-2}}^H\M}{\M^{H}\U_{\rm n_{\rm K}}\U_{\rm n_{\rm K}}^H\M}\\\ldots&\frac{\M^{H}\U_{\rm n_{\rm 1}}\U_{\rm n_{\rm 1}}^H\M}{\M^{H}\U_{\rm n_{\rm K}}\U_{\rm n_{\rm K}}^H\M}\frac{1}{\M^{H}\U_{\rm n_{\rm K}}\U_{\rm n_{\rm K}}^H\M},
%\end{align}
where $\U_K\triangleq[\mathbf{u}_{K+1},\ldots,\mathbf{u}_{N_{\rm RF}}]$ and $\M\triangleq\W_{\rm RX}^{\rm H}\P_{\rm RX}^{\rm H}\a_{\rm RX}(r,\theta,\varphi)$.
% \begin{align}\label{eq: spectral_peak}
%         \mathcal{M}(r,\theta,\varphi) \triangleq \left\|\a_{\rm RX}^{\rm H}(r,\theta,\varphi)\P_{\rm RX}\W_{\rm RX}\U_{n}\right\|^{-2}.
% \end{align}
A 3D search on this function's peaks will yield the estimates $(\widehat{r}_k,\widehat{\theta}_k,\widehat{\varphi}_k)$'s for all $K$ targets. %To reduce complexity, compressed sensing can be adopted~\cite{locrxris}.
%Performing a 3D spectral search over the range, elevation, and azimuth angle domain on \eqref{eq: spectral_peak} derives paired range, elevation, and azimuth angle estimates $(\widehat{r}_k,\widehat{\theta}_k,\widehat{\varphi}_k),\,\forall k$, of K radar target sources. Complexity can be reduced by setting ̂$\widehat{\varphi}_k = \frac{\pi}{2},\,\forall k$ since all radar targets are on the yz-plane. Note that further complexity reduction with RAnk-REduced (RARE) algorithm or compressed sensing method is possible, but is not within the scope of this paper.

\subsection{FD Holographic MIMO Optimization}
Our FD-enabled ISAC design objective is the joint maximization of the Signal-to-Noise-Ratios (SNRs) of the estimation of $(r_k,\theta_k,\varphi_k)$'s for all $K$ targets and of the DL for the $U$ UEs (i.e., the $U$ out of the $K$ targets). In mathematical terms, we focus on the following optimization problem:
\begin{align}
        \mathcal{OP}&:\nonumber\underset{\substack{\W_{\rm TX},\W_{\rm RX}\\ \V,\D}}{\max} \quad \widehat{ {\Gamma}}_{\rm R}+ \widehat{ {\Gamma}}_{\rm DL}\\
        &\nonumber\text{\text{s}.\text{t}.}\,
        \|[\W^{\rm H}_{\rm RX}\P^{\rm H}_{\rm RX}\H_{\rm SI}\P_{\rm TX}\W_{\rm TX}\V]_{(i,:)}\|^2\leq \gamma,\, \forall i,\\
        &\nonumber\,\quad \sum\limits_{u=1}^{U}\|\P_{\rm TX}\W_{\rm TX}\V_u\|^2 \leq P_{\rm max},\\
        &\,\quad w^{\rm TX}_{i,n} \in \mathcal{W},\, w^{\rm RX}_{i,n} \in \mathcal{W},\nonumber
\end{align}
where the SNRs $\widehat{{\Gamma}}_{\rm R}$ and $\widehat{{\Gamma}}_{\rm DL}$ are given by:
\begin{align}
        \widehat{{\Gamma}}_{\rm R} &\triangleq \left\|\W^{\rm H}_{\rm RX}\P^{\rm H}_{\rm RX}\widehat{\H}_{\rm R}\P_{\rm TX}\W_{\rm TX}\V\right\|^2\widehat{\Sigma}^{-1},\label{eq:SNR_R}\\
        \widehat{{\Gamma}}_{\rm DL} &\triangleq \sum\limits_{u=1}^{U}\Big(\|\widehat{\H}_{\rm DL,u}\P_{\rm TX}\W_{\rm TX}\V\|^2\sigma_u^{-2}\Big)
\end{align}
with $\widehat{\H}_{\rm R}$ and $\widehat{\H}_{\rm DL,u}$ $\forall$$u$ constructed using $(\widehat{r}_k,\widehat{\theta}_k,\widehat{\varphi}_k)$'s. In the $\mathcal{OP}$ formulation, the first constraint refers to the residual SI threshold $\gamma$ at the output of each $i$th microstrip of the RX DMA. In \eqref{eq:SNR_R}, $\widehat{\Sigma} \triangleq \|\widetilde{\H}_{\rm SI}\V\|^2 + \|\P_{\rm RX}\W_{\rm RX}\|^2\sigma^2$ represents the Interference-plus-Noise (IpN) after digital SI cancellation and analog combining at the baseband of the RX DMA. 

To solve the non-convex $\mathcal{OP}$, which has coupling variables, we employ an alternating optimization approach. By utilizing $(\widehat{r}_k,\widehat{\theta}_k,\widehat{\varphi}_k)$'s, the DL channels $\widehat{\H}_{{\rm DL},u}$ $\forall u$ are formulated using \eqref{eqn:DL_chan}, as well as the composite end-to-end channel $\widehat{\H}_{\rm R}$ using \eqref{eq:H_R}, but without the inclusion of $\beta_k$'s, which are unknown. To find the TX/RX DMA analog BF matrices $\W_{\rm TX}$ and $\W_{\rm RX}$, we first restrict their elements to the set $\mathcal{F}\in{e^{j\phi}|\phi\in\left[-\pi/2,\pi/2\right]}$ (e.g., DFT codebook) having constant amplitude and arbitrary phase values, and then solve the two following optimization problems sequentially via $1$D searches: 
%We use the estimated $(\widehat{r}k,\widehat{\theta}k,\widehat{\varphi}k),,\forall k$ to formulate the virtual radar channel $\widehat{\H}_{\rm R}$ and DL channels $\widehat{\H}_{{\rm DL},u},\forall u$, similar to \eqref{eqn:DL_chan}. Our objective is to find the TX/RX DMA weights $\W_{\rm TX}$ and $\W_{\rm RX}$ given the virtual radar and DL channel. However, due to the coupling between the phase and amplitude of $w^{\rm TX}_{i,n}$, finding the weights is challenging. Therefore, we propose to solve the optimization problems to find DMA weights from a set of constant amplitude and arbitrary phase values $\mathcal{F}\in{e^{j\phi}|\phi\in\left[-\pi/2,\pi/2\right]}$ (e.g., DFT codebook) as:
\begin{align*}
    \begin{array}{cc}
        \mathcal{OP}1:\underset{\widetilde{\W}_{\rm TX}}{\max} \,\, \|\widehat{\H}_{\rm R}\widetilde{\W}_{\rm TX}\|^2, & \!\!\mathcal{OP}2:\underset{\widetilde{\W}_{\rm RX}}{\max} \,\, \frac{\|\widetilde{\W}_{\rm RX}\widehat{\H}_{\rm R}\widetilde{\W}_{\rm TX}\|^2}{\|\widetilde{\W}_{\rm RX}\widehat{\H}_{\rm SI}\widetilde{\W}_{\rm TX}\|^2} \\
        \qquad\quad\text{\text{s}.\text{t}.}\qquad \widetilde{w}^{\rm TX}_{i,n}\in \mathcal{F}\quad  & \qquad\quad\text{\text{s}.\text{t}.}\qquad \widetilde{w}^{\rm RX}_{i,n}\in \mathcal{F}\quad
    \end{array}.
\end{align*}
Given $\widetilde{\W}_{\rm TX}$, $\widetilde{\W}_{\rm TX}$, and compensating for the signal propagation inside the microstrips, the final TX/RX DMA weights can be derived as follows:
\begin{align}\label{eq:weights}
    w^{\rm TX}_{i,n} \triangleq \frac{\jmath+\widetilde{w}^{\rm TX}_{i,n}e^{\jmath\rho_{i,n} \beta_{i}}}{2}, \,\, w^{\rm RX}_{i,n} \triangleq \frac{\jmath+\widetilde{w}^{\rm RX}_{i,n}e^{\jmath\rho_{i,n} \beta_{i}}}{2}.
\end{align}
Then, we find the multi-user digital BF matrix $\V$ employing block diagonalization, such that it maximizes the SNR of the DL UEs, while minimizing the inter-UE interference and suppressing the residual SI signal at the RX DMA's output below the required threshold. The proposed FD-enabled ISAC design solving $\mathcal{OP}$ is summarized in Algorithm~\ref{alg:the_opt}. It is noted that, in the algorithmic Step $1$, a random feasible $\W_{\rm RX}$ can be used when the ISAC system runs for the first time, or a matrix exploiting any prior knowledge. Then, the targets' parameters estimation can be performed with the optimized $\W_{\rm RX}$.
\begin{algorithm}[!t]
    \caption{FD Holographic MIMO ISAC}
    \label{alg:the_opt}
    \begin{algorithmic}[1]
        \renewcommand{\algorithmicrequire}{\textbf{Input:}}
        \renewcommand{\algorithmicensure}{\textbf{Output:}}
        \REQUIRE $\P_{\rm TX}$, $\P_{\rm RX}$, ${\H}_{\rm SI}$, $U$, and $P_{\rm max}$. %$(\widehat{r}_k,\widehat{\theta}_k,\widehat{\varphi}_k)$ $\forall k=1,\ldots,K$. 
        \ENSURE $\W_{\rm TX}$, $\W_{\rm RX}$, $\V$, and $\D$.
        \STATE Obtain $(\widehat{r}_k,\widehat{\theta}_k,\widehat{\varphi}_k)$ $\forall k=1,\ldots,K$ from \eqref{Spec_ROOT}'s peaks. 
        \STATE Set $\widehat{\H}_{\rm R} = \sum\limits_{k=1}^{K} \a_{\rm RX}(\widehat{r}_k,\widehat{\theta}_k,\widehat{\varphi}_k)\a_{\rm TX}^{\rm H}(\widehat{r}_k,\widehat{\theta}_k,\widehat{\varphi}_k)$\\ and construct $\widehat{\H}_{{\rm DL},u}$ $\forall$$u$ using \eqref{eqn:DL_chan}.
        \STATE Solve $\mathcal{OP}1$ and $\mathcal{OP}2$ to find $\widetilde{\W}_{\rm TX}$ and $\widetilde{\W}_{\rm RX}$.
        \STATE Compute $w^{\rm TX}_{i,n}$ and $w^{\rm RX}_{i,n}$ $\forall$$i,n$ using \eqref{eq:weights} and obtain $\W_{\rm TX}$ and $\W_{\rm RX}$.
        \STATE Set $\D = -(\W^{\rm H}_{\rm RX}\P^{\rm H}_{\rm RX}\H_{\rm SI}\P_{\rm TX}\W_{\rm TX})$ and derive $\B$ as the $N_{\rm RF}$ right-singular vectors of $-\D$.
        \IF{$U=1$}
                    \STATE Set $\G=\sqrt{P_{\rm max}}\E$ with $\E$ having the right-singular vectors of $\H_{{\rm eff},1}\!=\!\widehat{{\H}}_{\mathrm{DL},u}\P_{\rm TX}\W_{\rm TX}\B$.
                    \IF{$\|[{\W}^{\rm H}_{\rm RX}\P^{\rm H}_{\rm RX}{\H}_{\rm SI}\P_{\rm TX}{\W}_{\rm TX}\B\G]_{(i,:)}\|^2\leq \gamma,\,\forall i$}
    			\STATE Output $\V=\B\G$ and stop the algorithm.
    	        \ELSE
            			 \STATE Output that the FD holographic MIMO settings do not meet the residual SI constraints.
                    \ENDIF
        \ENDIF
        \FOR{$u=1,2,\ldots,U$}
        \FOR{$\alpha={N}_{\rm RF},N_{\rm RF}-1,\ldots,L$}
    		\STATE Set $\F\!=\![\B]_{(:,N_{\rm RF}-\alpha+1:N_{\rm RF})}$\\ and $\H_{{\rm eff},u}\!=\!\widehat{{\H}}_{\mathrm{DL},u}\P_{\rm TX}\W_{\rm TX}\F$.            
    		\STATE Set ${\K}_u\in \Compl^{N_{\rm RF}\times L}$ as the null space of the effective DL channel with the $u$th UE removed:\\ $\bar{\H}_{{\rm eff},u}\triangleq[\H_{{\rm eff},1},\ldots,\H_{{\rm eff},u-1},\H_{{\rm eff},u+1},\ldots,\H_{{\rm eff},U}]$.
            \STATE Set $\bar{\E}_u\in \Compl^{N_{\rm RF}\times (L-1)}$ as the right-singular vectors of ${\K}_u$.
    		\STATE Set ${\E}_u$ as the right singular vectors of $\H_{{\rm eff},u}\bar{\E}_u$.
    		\STATE Set $\G_u = \sqrt{{P_{\rm max}/U}}\bar{\E}_u{\E}_u$ as the optimum block-diagonalized precoder for the $u$th UE.
        \ENDFOR
           \STATE Construct digital precoder $\G=[\G_1,\dots,\G_U]$.
    		\IF{$\|[{\W}^{\rm H}_{\rm RX}\P^{\rm H}_{\rm RX}{\H}_{\rm SI}\P_{\rm TX}{\W}_{\rm TX}\F\G]_{(i,:)}\|^2\leq \gamma,\,\forall i$}
    			 \STATE Output $\V=\F\G$ and stop the algorithm.
    	    \ELSE
            			 \STATE Output that the FD holographic MIMO settings do not meet the residual SI constraints.
    		\ENDIF
    	\ENDFOR
    \end{algorithmic}
\end{algorithm}

\section{Numerical Results and Discussion}\label{sec: num}
In this section, we numerically evaluate the ISAC performance of the proposed FD holographic MIMO framework, when operating in the sub-THz frequency band and in the near-field regime. We have simulated a scenario including $K=3$ sensing targets, with $U=2$ being the UEs each with $L=2$ antennas, whose direction and range need to be estimated. The proposed FD holographic MIMO node was assumed to deploy TX/RX DMAs, each consisting of $N_{\rm RF} = \{4,5,6\}$ microstrips with each having $N_{\rm E}=512$ metamaterials. The microstrips at both DMAs were placed along the $x$-axis, as shown in Fig.~\ref{fig: FD_ISAC}, with inter-microstrip distance $d_{\rm RF}=\lambda/2$. The inter-element distance within each microstrip was set as $d_{\rm E}=\lambda/5$ and the separation between the TX and RX DMAs was chosen as $d_{pl}=2d_{\rm P}=0.04$ meters. The central frequency of the proposed ISAC system was $120$ GHz covering a bandwidth $B=150$ KHz. Our Monte Carlo runs for the performance evaluation were designed as follows: we have used $T=200$ transmission time slots for communications and sensing per UE location, and placed the UE randomly in an environment with a fixed azimuth at $90^{\circ}$, elevation lying in the set $[0^{\circ},90^{\circ}]$, and a range between $1$ and $25$ meters (propagation within the Fresnel region). Finally, the noise variances $\sigma^2$ and $\sigma_1^2$ in dB were set to $-174 + 10\log_{10}(B)$, $\beta_1$ appearing in \eqref{eq:H_R} was chosen randomly with unit amplitude, and we have used a $10$-bit beam codebook $\mathcal{F}$ for both the TX/RX DMA analog BF matrices $\W_{\rm TX}$ and $\W_{\rm RX}$ in $\mathcal{OP}1$.
\begin{figure*}[!t]
  \begin{subfigure}[t]{0.45\textwidth}
  \centering
    \includegraphics[width=\textwidth]{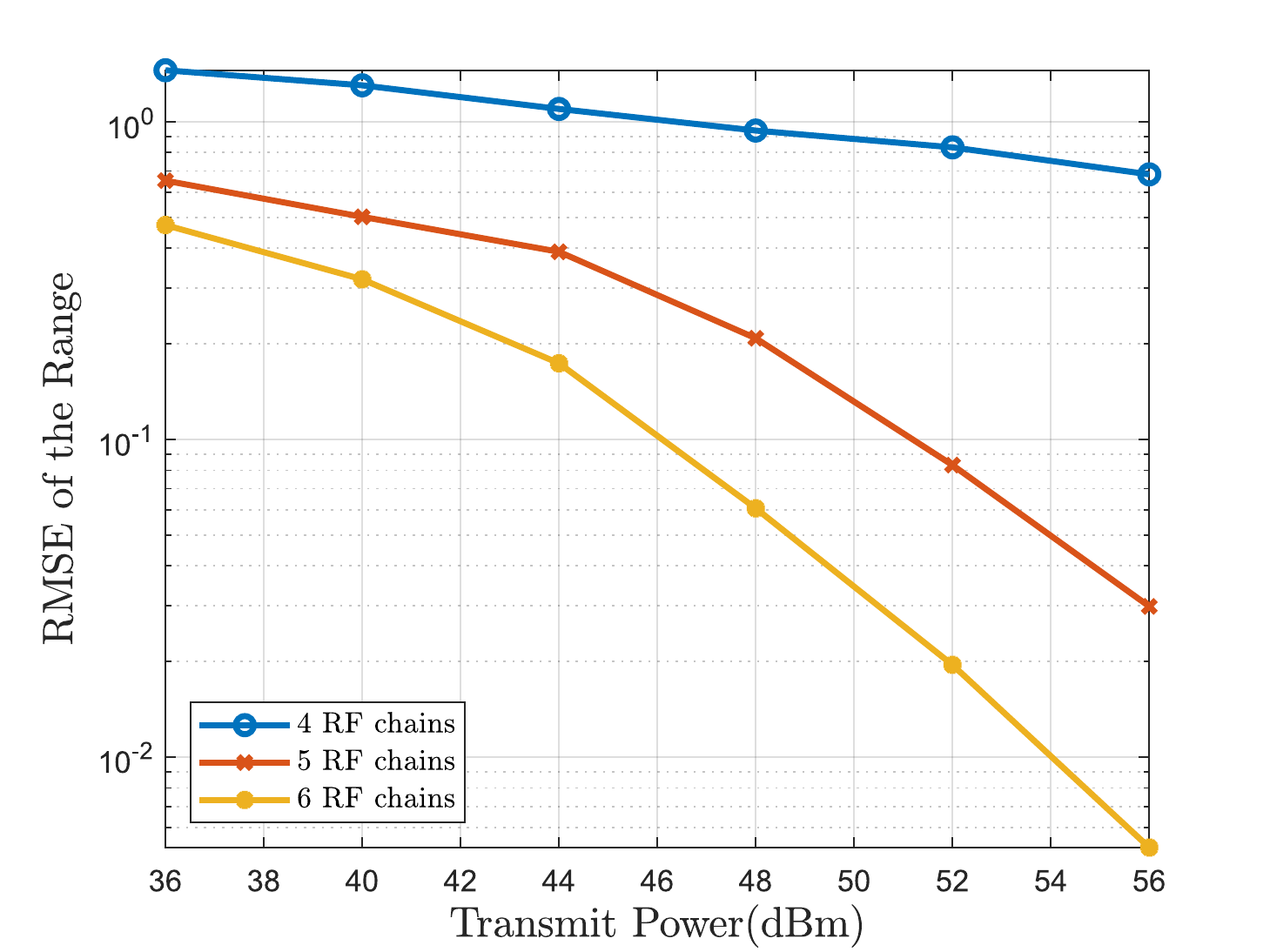}
    \caption{Range estimation.}
    \label{fig:Range_opt}
  \end{subfigure}\hfill
  \begin{subfigure}[t]{0.45\textwidth}
  \centering
    \includegraphics[width=\textwidth]{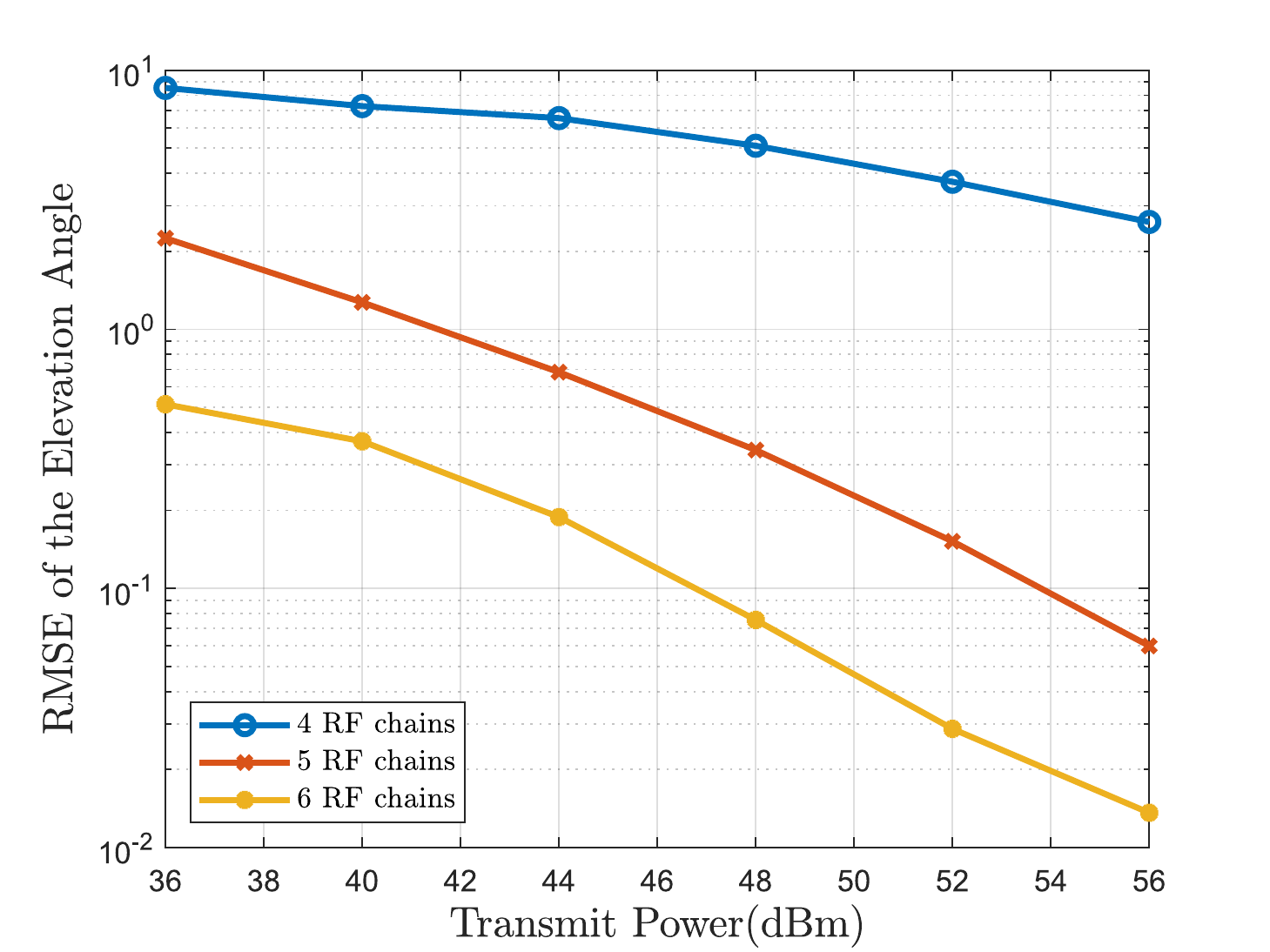}
    \caption{Estimation of the elevation angle.}
    \label{fig:Theta_opt}
  \end{subfigure}
  \caption{\small{Average sensing estimation performance for $U=2$ UEs, each with $L=2$ antenna elements, versus the transmit power $P_{\rm max}$ in dBm, considering an FD holographic MIMO node with $N_{\rm RF} = \{4,5,6\}$} TX/RX microstrips each with $N_{\rm E}=512$ metamaterials.}\vspace{-0.4cm}
  \label{fig:Estimation_vs_P_T}
\end{figure*}

\begin{figure}[!t]
\centering
\includegraphics[width=0.92\columnwidth]{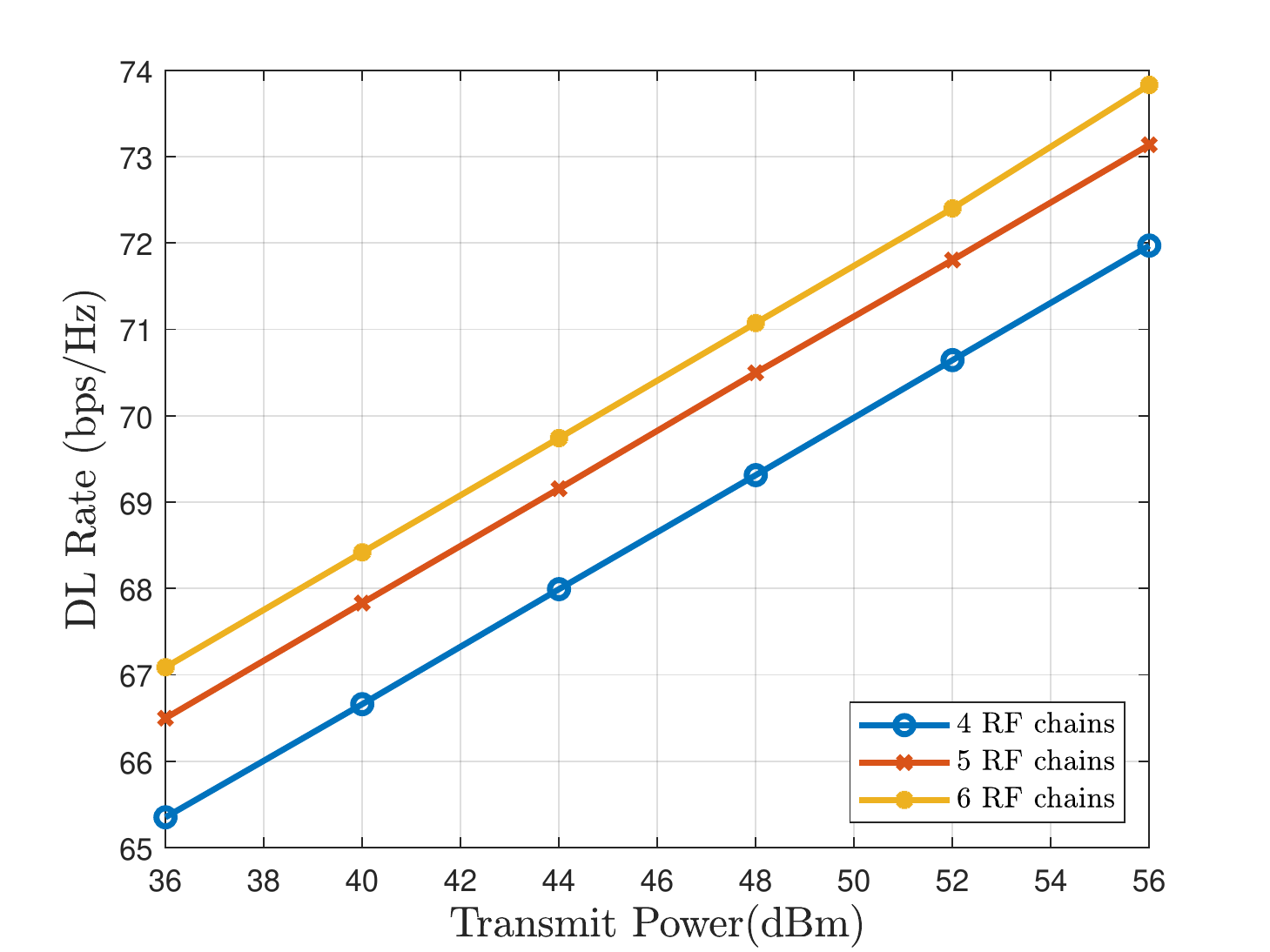}\vspace{-0.2cm}
\caption{\small{Achievable DL rate versus the transmit power $P_{\rm max}$ in dBm for the parameters' setting in Fig.~\ref{fig:Estimation_vs_P_T}.}
}\vspace{-0.4cm}
\label{fig:DL}
\end{figure}

In Figs.~\ref{fig:Estimation_vs_P_T} and~\ref{fig:DL}, simulation results for the proposed FD-enabled ISAC scheme detailed in Algorithm~\ref{alg:the_opt} are illustrated for different total transmit power levels $P_{\rm max}$ in dBm, which are typical for sub-THz wireless communications. In particular, Figs.~\ref{fig:Range_opt} and~\ref{fig:Theta_opt} demonstrate the Root Mean Square Error (RMSE) of the estimations for the range and the elevation angle, respectively, whereas Fig.~\ref{fig:DL} depicts the achievable DL rate performance in bps/Hz. As expected, all performance metrics improve with increasing SNR values, and it is showcased that, increasing the number of microstrips (consequently, the number of TX/RX RF chains) improves both estimation performance and the DL rate. The latter is reasonable since more spatial sampling improves the target parameter estimation, which in turn, improves DL channel estimation. This, in conjunction with, the larger BF gain, results in rate boosting.   

\section{Conclusion}
In this paper, we presented an FD holographic MIMO ISAC system, operating in THz frequencies and in the near-field regime, capable to offer simultaneous multi-user DL communication and sensing of targets being randomly distributed within its coverage area. A novel optimization framework for the joint design of the TX/RX DMAs' analog BF matrices, the TX digital BF matrix, and the digital SI cancellation matrix was devised. Our numerical results showcased the ISAC potential of the proposed scheme for various system parameters.

\bibliographystyle{IEEEtran}
\bibliography{IEEEabrv,ms}

\begin{thebibliography}{10}
\providecommand{\url}[1]{#1}
\csname url@samestyle\endcsname
\providecommand{\newblock}{\relax}
\providecommand{\bibinfo}[2]{#2}
\providecommand{\BIBentrySTDinterwordspacing}{\spaceskip=0pt\relax}
\providecommand{\BIBentryALTinterwordstretchfactor}{4}
\providecommand{\BIBentryALTinterwordspacing}{\spaceskip=\fontdimen2\font plus
\BIBentryALTinterwordstretchfactor\fontdimen3\font minus
  \fontdimen4\font\relax}
\providecommand{\BIBforeignlanguage}[2]{{%
\expandafter\ifx\csname l@#1\endcsname\relax
\typeout{** WARNING: IEEEtran.bst: No hyphenation pattern has been}%
\typeout{** loaded for the language `#1'. Using the pattern for}%
\typeout{** the default language instead.}%
\else
\language=\csname l@#1\endcsname
\fi
#2}}
\providecommand{\BIBdecl}{\relax}
\BIBdecl

\bibitem{liu2022integrated}
F.~Liu \emph{et~al.}, ``Integrated sensing and communications: Towards
  dual-functional wireless networks for {6G} and beyond,'' \emph{IEEE J. Sel.
  Areas Commun.}, 2022.

\bibitem{mishra2019toward}
K.~V. Mishra \emph{et~al.}, ``Toward millimeter-wave joint radar
  communications: A signal processing perspective,'' \emph{IEEE Signal Process.
  Mag.}, vol.~36, no.~5, pp. 100--114, Sep. 2019.

\bibitem{sabharwal2014band}
A.~Sabharwal \emph{et~al.}, ``In-band full-duplex wireless: Challenges and
  opportunities,'' \emph{IEEE J. Sel. Areas Commun.}, vol.~32, no.~9, pp.
  1637--1652, Sep. 2014.

\bibitem{alexandropoulos2017joint}
G.~C. Alexandropoulos and M.~Duarte, ``Joint design of multi-tap analog
  cancellation and digital beamforming for reduced complexity full duplex
  {MIMO} systems,'' in \emph{Proc. IEEE ICC}, Paris, France, May 2017.

\bibitem{FD_MIMO_VTM2022}
G.~C. Alexandropoulos \emph{et~al.}, ``Full duplex massive {MIMO}
  architectures: {R}ecent advances, applications, and future directions,''
  \emph{IEEE Veh. Technol. Mag.}, vol.~17, no.~4, pp. 83--91, Dec. 2022.

\bibitem{Vishwanath_2020}
I.~P. Roberts \emph{et~al.}, ``Equipping millimeter-wave full-duplex with
  analog self-interference cancellation,'' in \emph{Proc. {IEEE ICC}}, Ireland,
  Jun. 2020.

\bibitem{alexandropoulos2020full}
G.~C. Alexandropoulos \emph{et~al.}, ``Full duplex hybrid {A/D} beamforming
  with reduced complexity multi-tap analog cancellation,'' in \emph{Proc. {IEEE
  SPAWC}}, Atlanta, USA, May 2020.

\bibitem{barneto2019full}
C.~B. Barneto \emph{et~al.}, ``Full-duplex {OFDM} radar with {LTE} and {5G NR}
  waveforms: Challenges, solutions, and measurements,'' \emph{IEEE Trans.
  Microw. Theory Techn.}, vol.~67, no.~10, pp. 4042--4054, Aug. 2019.

\bibitem{liyanaarachchi2021optimized}
S.~D. Liyanaarachchi \emph{et~al.}, ``Optimized waveforms for {5G--6G}
  communication with sensing: Theory, simulations and experiments,'' \emph{IEEE
  Trans. Wireless Commun.}, Jun. 2021.

\bibitem{barneto2020beamforming}
C.~B. Barneto \emph{et~al.}, ``Beamforming and waveform optimization for
  {OFDM}-based joint communications and sensing at mm-waves,'' in \emph{Proc.
  IEEE ASILOMAR}, Pacific Grove, USA, Nov. 2020, pp. 895--899.

\bibitem{Islam_2022_ISAC}
M.~A. Islam \emph{et~al.}, ``Integrated sensing and communication with
  millimeter wave full duplex hybrid beamforming,'' in \emph{Proc. IEEE ICC},
  Seoul, South Korea, May. 2022.

\bibitem{Atiq_ISAC_2022}
------, ``Simultaneous multi-user {MIMO} communications and multi-target
  tracking with full duplex radios,'' in \emph{Proc. {IEEE GLOBECOM}}, Rio de
  Janeiro, Brazil, Dec. 2022.

\bibitem{Shlezinger2021Dynamic}
N.~Shlezinger \emph{et~al.}, ``Dynamic metasurface antennas for {6G} extreme
  massive {MIMO} communications,'' \emph{{IEEE} Wireless Commun.}, vol.~28,
  no.~2, pp. 106--113, Apr. 2021.

\bibitem{HMIMO_survey}
T.~Gong \emph{et~al.}, ``Holographic {MIMO} communications: {T}heoretical
  foundations, enabling technologies, and future directions,'' \emph{arXiv
  preprint arXiv:2212.01257}, 2022.

\bibitem{Xu_DMA_2022}
J.~Xu \emph{et~al.}, ``Near-field wideband extremely large-scale {MIMO}
  transmission with holographic metasurface antennas,'' \emph{arXiv preprint
  arXiv:2205.02533}, 2022.

\end{thebibliography}
\end{document}